\def\gsim{\mathrel{\rlap{\lower.6ex\hbox{$\sim$}}\raise.35ex\hbox{$>$}}}
\def\lsim{\mathrel{\rlap{\lower.6ex\hbox{$\sim$}}\raise.35ex\hbox{$<$}}}
\documentstyle[preprint,aps,eqsecnum]{revtex}
\begin{document}
\preprint{December 1993}
\title
{Spin gap fixed points in the double chain problem}

\author{D.V.Khveshchenko $^{a,b}$ and T.M.Rice $^{c}$}
\address
{$^a$ Department of Physics Princeton University,\\
Princeton, NJ 08544, USA\\
$^b$ Landau Institute for Theoretical Physics,\\
2,st.Kosygin, 117940, Moscow, Russia\\
$^c$ Theoretische Physik ETH-H\"{o}nggerberg,\\
 CH-8093, Z\"{u}rich, Switzerland}

\maketitle

\begin{abstract}
\noindent
Applying the bosonization procedure
to weakly  coupled Hubbard chains
 we discuss the fixed points of the renormalization group procedure  where
all spin excitations are gapful and a singlet pairing becomes the
dominant instability.
\end{abstract}
\pagebreak

\section{Introduction}
Recently the field
of non-Landau Fermi liquid states in various quasi-one dimensional systems
has been very active.
Although basic properties of purely one dimensional systems (chains) are
 quite well known by now,
it still remains to be understood how these change under coupling between
chains.
In particular, if an infinite number of chains is coupled to form
a two dimensional array then some kind of a dimensional crossover occurs.
Despite numerous intensive studies of these questions, it still
remains open how
the non-Landau Fermi liquid
one dimensional features evolve to isotropic two dimensional
behavior.

On the other  hand one may expect new physics in a system with a finite
number of coupled chains which may exhibit an unusual amalgamation
of both one- and two-dimensional
features.
Besides a  general theoretical interest, such systems are also attractive
because they  can be found in real materials.
Recently it was pointed out that some substances such as
$Sr_2Cu_4O_6$ provide a physical realization
of weakly coupled double chains \cite{RGS}.
Moreover higher stoichiometric compounds
in the series $Sr_{n-1}Cu_{n+1}O_{2n}$ present
examples of coupled $N$-chain ladders.

This expectation is reinforced by the behavior of $S=1/2$ Heisenberg
multichain or ladder systems. Whereas the single spin chain shows quasi-long
range order with gapless spinon excitations, the double chain
system shows spin liquid behavior \cite{H},\cite{DRS},
\cite{BDRS} with strictly short range order
and finite gap in the spin excitation spectrum.
This contrast in behavior has led to the conjecture that a lightly doped
double-chain system should preserve the spin gap and become superconducting
\cite{RGS}. Further in view of the insensitivity of the spin liquid state
to the ratio of interchain to intrachain coupling \cite{BDRS},
one expects that it is robust and should occur also for a Hubbard model
even in the weak coupling regime.
This  expectation is strongly supported by a recent numerical
study \cite{NWS}.
 This is our motivation to examine
the renormalization group (RG) theory of weakly coupled Hubbard double chains
and  to look for a spin liquid fixed  point.

Recent weak coupling RG
studies of the double-chain Hubbard model
\cite{FPT}
revealed some strong coupling fixed points characterized  by enhanced
singlet pairing.
However the analysis performed in \cite{FPT}
 is essentially restricted by the case of weak interactions between the
fermions.
These authors did not examine the half-filled case with weak interchain
hopping  where the Umklapp processes on the individual chains become relevant.

 In the present
paper we undertake an attempt to construct a description of
the spin gap fixed point by using
a bosonic representation. This is expected to be an adequate tool to
 demonstrate a development of the strong coupling
regime in both cases of weakly coupled Hubbard chains and of a
strongly correlated double chain $t-J$ model.
The latter case will be considered in a later publication  \cite{Khvesh}.

To clarify the essence of the double chain physics in the presence
of strong correlations
 it is worthwhile to begin with a review of well known properties of
the single chain Hubbard model.

Away from half filling  the model
can be only found in the so-called Tomonaga- Luttinger (TL)
regime which corresponds to both gapless spin and charge excitations
\cite{Sch}.
It is customary to describe the TL  behavior in terms of spin and charge
correlation exponents $K_{s}$ and
$K_{c}$.

The spin exponent $K_{s}$ equals to unity everywhere in the TL regime while
$K_{c}$
gradually increases from the value  $K_{c}=1/2$ at values of the onsite
repulsion $U=\infty$
 and any electron density $\rho\neq 1$
(as well as at $\rho\rightarrow 1$
and arbitrary $U/t$ ($t$: intrachain hopping))
as $U$ increases or $\rho$ gets smaller.

In the regime of strong correlation at $\rho$ close to unity
one can argue that gapless spin fluctuations drive  the
coupling constants  of the charge
sector
to the repulsive region  $(K_{c}<1)$.
In this case we can expect a change of behavior when the exchange coupling
between chains is introduced. The spin gap, which we have argued is a robust
feature of the single rung ladder at $\rho=1$, acts to cutoff the spin
fluctuation spectrum at low energies so that these may not renormalize
the charge couplings significantly. As a result there can be an effective
attraction at $\rho \lsim 1$ without a threshold value of the ratio $U/t$.
This will manifest itself in a finite spin gap also at $\rho\lsim 1$
and  a scaling to the Luther-Emery line rather than the Tomonaga-Luttinger
line. Such behavior has been reported for single chains models with a spin
gap caused by frustrated or modulated exchange couplings on the single chain
\cite{I},\cite{OLR}. In the present case of a single rung ladder or double
chain, we will choose the regime with the interchain hopping $t_\perp$ weak
$(t_\perp<<U)$ so that as $\rho\rightarrow 1$ real interchain kinetic energy
processes characterized by $t_\perp$ will scale to zero. In this limit
the induced interchain exchange processes will remain finite and dominate.

A mean field analysis \cite{SRZ} of the $t-J$ model predicts that
the spin gap remains upon doping and the RVB
state at $\rho=1$ evolves into a superconductor
with approximate d-wave symmetry. Noack ${\it et}$ ${\it al}$ \cite{NWS} found
a similar behavior in their numerical studies of moderately coupled
Hubbard ladders.

In this paper we investigate the case of a weakly coupled Hubbard ladder
with $\rho\lsim 1$ and $t_\perp <U<t$
using RG methods and look for a strong coupling fixed point with the same
characteristics.

\section{Bosonized weak coupling limit of the double-chain problem}
To get the first insight into the problem we start with a conventional
bosonization of the small $U/t$ Hubbard model on two weakly coupled chains.
\begin{eqnarray}
H=t\sum( u^{\dagger}_{i\sigma}
u_{i+1,\sigma}+d^{\dagger}_{i\sigma}d_{i+1,\sigma})+
t_\perp \sum
(d^{\dagger}_{i\sigma}
u_{i\sigma}+u^{\dagger}_{i\sigma}d_{i\sigma}) +\nonumber\\
+U \sum
(u^{\dagger}_{i\sigma}u_{i\sigma} u^{\dagger}_{i,-\sigma}u_{i,-\sigma}+
d^{\dagger}_{i\sigma}d_{i\sigma} d^{\dagger}_{i,-\sigma}d_{i,-\sigma})
\end{eqnarray}
Here $u_{i\sigma}$
and $d_{i\sigma}$ denote fermions on upper ("u") and lower ("d")
chains.

Apparently, at $U<<t_\perp ,t$
the interaction term has to be treated as a small
perturbation to the rest of the Hamiltonian (2.1) and
the bare transverse hopping leads to the formation of two (bonding (B)
and antibonding (A)) bands: $(A,B)={1\over{\sqrt{2}}}(u\mp d)$.
 Thus at $U<<t_{\perp}$ a proper starting point is provided
by the two-band model
which was previously studied in the framework of a general weak coupling
 "g-ology"\cite{VZ},\cite{PS}.
The analysis carried out in \cite{FPT} was also based on the two-band picture.

However, if  the opposite condition $t_{\perp}<<U$ is satisfied then
the effect of band splitting is completely suppressed due to the
requirement to avoid a double on-site occupancy. This behavior persists
down to the quarter filling ( $\rho =1/2$).
In the framework of the RG approach this
 phenomenon manifests itself as a vanishing of the renormalized $t_\perp$.
 In view of this we suppose that in the case $t_\perp<<U$
one has to start from the picture of two degenerate bands
 to implement correctly the fermion correlations.

The  preceding RG analysis of  the
general two-band model in absence of Umklapp processes
\cite{VZ},\cite{PS}
already shows many  technical complexities.
For this reason the results of these studies
are not simply physically transparent. Moreover it turns out that all
nontrivial fixed points are located far
 in the strong coupling regime where the lowest
orders RG calculations cease to be valid.
So one might expect that a
more informative investigation can be done on the basis of a
bosonic representation which is usually capable of yield of
a correct evolution
toward strong coupling and even
to provide an asymptotically exact solution of the Luther-Emery type
\cite{ME}. Recently the method of bosonization was applied to the double
chain problem in the context of a special model which includes
 only forward scattering
\cite{FL}. This analysis \cite{FL} led to the prediction that
coupled chains provide a proper basis for the occurence of singlet pairing.

 In this paper we will perform a more general analysis
than that of \cite{FL} to see whether the above statement holds
for a wider class of models.

To proceed with a bosonic representation
 we introduce a conventional set
of bosonic fields
$\phi_c^{f}$, $\phi_s^{f}$
where the "flavor index" $f$ has  one of two values $u$ or $d$.
These fields describe fluctuations of charge $(c)$ and
spin $(s)$ densities respectively. In the continuum limit
the fermion operators can be written in terms of these variables as follows
\begin{eqnarray}
\Psi^{f}_{\sigma}(x)\sim \sum_{\mu}\exp(i\mu k_{F}x+
{i\over{\sqrt 2}}(\mu\phi^{f}_{c}+\theta^{f}_{c}+\mu\sigma\phi^{f}_{s}+
\sigma\theta^{f}_{s}))
\end{eqnarray}
where $\theta^{f}_{c,s}$ are the dual fields
to the $\phi^{f}_{c,s}$ $(\partial_{\mu}\theta^{f}_{c,s}=
\epsilon_{\mu\nu}\partial_{\nu}\phi^{f}_{c,s})$.

Applying the  formula (2.2) and introducing the linear combinations
$\phi_{c,s}^{\pm}={i\over{\sqrt 2}}(\phi^{u}_{c,s}\pm\phi^{d}_{c,s}),
\theta_{c,s}^{\pm}={i\over{\sqrt 2}}(\theta^{u}_{c,s}\pm\theta^{d}_{c,s})$
corresponding to total
 $("+")$ and relative $("-")$ charge  or spin
density fluctuations,
one can readily obtain the bosonic form of the Hamiltonian (2.1)
\begin{eqnarray}
H_B={1\over 2}\sum_{\pm}(v_{c}K_{c}^{\pm}(\partial\theta_{c}^{\pm})^2
+{v_{c}\over K_{c}^{\pm}}
(\partial\phi_{c}^{\pm})^2+v_{s}K_{s}^{\pm}(\partial\theta_{s}^{\pm})^2
+{v_{s}\over K_{s}^{\pm}}
(\partial\phi_{s}^{\pm})^2)+\nonumber\\
+t_\perp (\cos \phi^{-}_{c}
\cos \phi^{-}_{s}+\cos (\phi^{+}_{c}+\delta x)\cos \phi^{+}_{s})
\cos\theta^{-}_{c}\cos\theta^{-}_{s}+\nonumber\\
g_{BS}\cos 2\phi^{+}_{s}
\cos 2\phi^{-}_{s}+g_{U}\cos (2\phi^{+}_{c}+2\delta x)\cos 2\phi^{-}_{c}
\end{eqnarray}
where last two
 cosine term represents spin backscattering and
Umklapp processes respectively.
Each of these terms is, in fact, a sum of two
contributions $\cos 2{\sqrt 2}\phi_{c,s}^{f}$ coming from  $u$ and $d$
species. The Umklapp term
 becomes relevant when the doping $\delta (={\pi\over 2}-k_{F})$ vanishes.
As usual, the bare
values of the correlation exponents $K_{c,s}^{\pm}=\frac{v_{F}}
{v_{c,s}}$
  can be
changed by shortwavelength renormalizations. Neglecting these corrections
we obtain that the bare correlation exponents governing the charge dynamics
$K_c^{\pm}(=(1+{U\over {\pi t}})^{-1/2})$
 are smaller than unity while the spin
exponents
$K_s^{\pm}(=(1+{U\over {\pi t}})^{1/2})$ are opposite
and $K_s^{\pm}>1$. In addition, the bare values of the
amplitudes $g_{BS}$ and $g_U$ are equal to ${U\over {\pi t}}$.

To perform a renormalization procedure we divide up all variables
on slow and fast components and then integrate out the fast variables.
Using the bare values of correlation exponents one can estimate scaling
dimensions of various terms in (2.3) according to the conventional formula
\cite{C}($\gamma$ and $\gamma^\prime$ are arbitrary):
\begin{equation}
\Delta(\cos\gamma\phi_{c,s}^{\pm}\cos\gamma^{\prime}
\theta_{c,s}^{\pm})
={1\over 4}(\gamma^{2} K_{c,s}^{\pm}+
{{\gamma^{\prime}}^2\over {K_{c,s}^{\pm}}})
\end{equation}
The Hamiltonian (2.3) has to be supplemented by
extra terms which are generated in
the course of renormalization.
Indeed, performing an
expansion of the partition function $Z=Tr\exp(-\beta H_B)$
in $t_\perp$ one
immediately observes that the
interchain hopping produces the
following relevant terms (new couplings $g_i$ should not be confused with the
traditional g-ological notations):
\begin{eqnarray}
\Delta H=g_{1} \cos2\phi_{c}^{-}\cos 2\theta_{s}^{-}+
g_{2} \cos2\theta_{c}^{-}\cos 2\phi_{s}^{-}+
g_{3} \cos2\theta_{c}^{-}\cos 2\phi_{s}^{+}+\nonumber\\
g_{4} \cos(2\phi_{c}^{+}+2\delta x)\cos 2\theta_{s}^{-}+
g_{5} \cos2\phi_{c}^{-}\cos 2\phi_{s}^{-}+
g_{6} \cos2\phi_{c}^{+}\cos 2\theta_{s}^{-}+\nonumber\\
g_{7} \cos(2\phi_{c}^{+}+2\delta x)\cos 2\theta_{c}^{-}+
g_{8} \cos(2\phi_{c}^{+}+2\delta x)\cos 2\phi_{s}^{+}+
g_{9} \cos2\theta_{c}^{-}\cos 2\theta_{s}^{-}
\end{eqnarray}
All these terms have scaling dimensions not greater than two and
result from the second order perturbation corrections to the single chain
Hamiltonian
$\Delta H\sim
t_{\perp}^{2}<(\sum_{i}u^{\dagger}_{i}d_{i}+d^{\dagger}_{i}u_{i})^2>$.

Physically these terms  correspond to processes of
 coherent interchain  particle-hole and
particle-particle hopping triggered by the single particle one.
The crucial importance of these processes was previously pointed out
by many authors (see, for instance, \cite{S},\cite{BC}).

In the second order in $t_\perp$ the RG equations derived by the use of the
method of \cite{GS}
have the following form $(\xi=\ln x)$:
\begin{equation}
\frac{d g_{1}}{d\xi}
=(2-K_{c}^{-}-{1\over K_{s}^{-}})g_1 +{t_{\perp}^2 \over 2}
(K_{c}^{-}+{1\over K_{s}^{-}}-K_{s}^{-}-{1\over K_{c}^{-}})-g_{4}g_{U}
\end{equation}
\begin{equation}
{{d g_2}\over {d\xi}}
=(2-K_{s}^{-}-{1\over K_{c}^{-}})g_2 +{t_{\perp}^2 \over 2}
(K_{s}^{-}+{1\over K_{c}^{-}}-K_{c}^{-}-{1\over K_{s}^{-}})-g_3 g_{BS}
\end{equation}
\begin{equation}
{{dg_{3}}\over {d\xi}}
=(2-K_{s}^{+}-{1\over K_{c}^{-}})g_3 +{t_{\perp}^2 \over 2}
(K_{s}^{+}+{1\over K_{c}^{-}}-K_{c}^{-}-{1\over K_{s}^{+}})-g_2 g_{BS}
\end{equation}
\begin{equation}
{{dg_{4}}\over {d\xi}}=
(2-K_{c}^{+}-{1\over K_{s}^{-}})g_4 +{t_{\perp}^2 \over 2}
(K_{c}^{+}+{1\over K_{s}^{-}}-K_{s}^{+}-{1\over K_{c}^{-}})-g_1 g_U
\end{equation}
\begin{equation}
{{dg_{5}}\over {d\xi}}=(2-K_{c}^{-}-{K_{s}^{-}})g_5 +{t_{\perp}^2 \over 2}
(K_{c}^{-}+{K_{s}^{-}}-{1\over K_{s}^{-}}-{1\over K_{c}^{-}})
\end{equation}
\begin{equation}
{{dg_{6}}\over {d\xi}}=
(2-K_{s}^{+}-{1\over K_{s}^{-}})g_6 +{t_{\perp}^2 \over 2}
(K_{s}^{+}+{1\over K_{s}^{-}}-K_{c}^{+}-{1\over K_{c}^{-}})-g_3 g_9-g_4 g_8
\end{equation}
\begin{equation}
{{dg_{7}}\over {d\xi}}=
(2-K_{c}^{+}-{1\over K_{c}^{-}})g_7 +{t_{\perp}^2 \over 2}
(K_{c}^{+}+{1\over K_{c}^{-}}-K_{s}^{+}-{1\over K_{s}^{-}})-g_4 g_9
 \end{equation}.
\begin{equation}
{{dg_{8}}\over {d\xi}}=
(2-K_{c}^{+}-{1\over K_{s}^{+}})g_8 +{t_{\perp}^2 \over 2}
(K_{c}^{+}-{1\over K_{s}^{-}}+K_{s}^{+}-
{1\over K_{c}^{-}})-g_3 g_7 \end{equation}
\begin{equation}
{{dg_{9}}\over {d\xi}}=
(2-{1\over K_{c}^{-}}-{1\over K_{s}^{-}})g_9 +{t_{\perp}^2 \over 2}
(-K_{c}^{-}+{2\over K_{s}^{-}}-
K_{s}^{-}+{2\over K_{c}^{-}}-K_{s}^{+}-K_{s}^{+})-g_4 g_7-g_3 g_6
\end{equation}
\begin{equation}
{{dg_{BS}}\over {d\xi}}=(2-K_{s}^{-}-{K_{s}^{+}})g_{BS} -g_2 g_3
\end{equation}
\begin{equation}
{{dg_{U}}\over {d\xi}}=(2-K_{c}^{-}-{K_{c}^{+}})g_U -g_1 g_4 \end{equation}

\begin{equation}
{{d\log K_{c}^{-}}\over {d\xi}}={1\over 2}(-K_{c}^{-}(g_{1}^2+g_{5}^2+g_{U}^2)+
{1\over K_{c}^{-}}(g_{2}^2 +
g_{3}^2+g_{7}^2+g_{9}^2))
\end{equation}
\begin{equation}
{{d\log K_{s}^{-}}\over {d\xi}}={1\over 2}(-K_{s}^{-}(g_{2}^2+g_{5}^2+g_{BS}^2)
+{1\over K_{s}^{-}}(g_{1}^2+g_{4}^2+g_{6}^2+g_{9}^2))
\end{equation}
\begin{equation}
{{d\log K_{c}^{+}}\over {d\xi}}=-{1\over 2}K_{c}^{+}(g_{4}^2+
g_{7}^2+g_{8}^2+g_{U}^2)
\end{equation}
\begin{equation}
{{d\log K_{s}^{+}}\over {d\xi}}=-{1\over 2}K_{s}^{+}(g_{3}^2+
g_{6}^2+g_{8}^2+g_{BS}^2)
\end{equation}
\begin{equation}
{{d\log t_{\perp}}\over {d\xi}}=2-{1\over 4}(K_{c}^{-}+\frac{1}{K_{c}^{-}}
+K_{s}^{-}+\frac{1}{K_{s}^{-}})
\end{equation}
In addition, there are two equations describing evolutions of velocities
$v_{c,s}$ but one can always include these corrections into the definition of
the correlation
exponents.

 In comparison with the equations
obtained in \cite{VZ},\cite{PS} our RG equations (2.6-2.21) are
already written in terms of physically
relevant combinations  of original "g-ological"
 couplings, so one could hope that this description might appear to be
 more transparent. As we shall show the above  system consistently
demonstrates
a development of the strong coupling regime in rather general conditions,
so  we don't account next-to-leading order
 corrections which can be only necessary if one
discusses fixed points at finite coupling.

First, in the case of spinless
fermions away from half filling the only relevant
couplings are $g_1, g_2$ and $K_{c}^{-}$ and the equations (2.6-2.21) reduce
to those found previously  \cite{KLN},\cite{Y},\cite{NLK}.

However in the physically relevant case of spin one half fermions
 away from $\rho=1$  one can only neglect couplings $g_4, g_7, g_8$ and $g_U$
 associated with Umklapp processes and then the number of residual couplings
is large (10) and coincides with that
found in \cite{FPT}.

The fact that eq.(2.6-2.21) originate from the repulsive Hubbard model
simplifies their analysis
significantly. To see that one can choose a two-step renormalization procedure
to that of Ref.\cite{NLK} and integrate the above equations
first up to the scale $\xi_0=\ln {t\over t_\perp}$ where the renormalized
amplitude of single particle hopping $t_\perp (\xi)$ becomes of order unity
(and stops). It can be easily seen
that at $\xi\sim \xi_0$
one can still neglect renormalizations of the correlation exponents
$K^{\pm}_{c}=1-{z^{\pm}_{c}\over 2}$ and
$K^{\pm}_{s}=1-{z^{\pm}_{s}\over 2}$ from their bare
values corresponding to $z^{\pm}_{c}=-z^{\pm}_{s}=\lambda={U\over {\pi t}}$.

By straightforward generalization of the analysis of \cite{Y},\cite{NLK}
one  can obtain  the evolution of couplings $g_i (\xi)$ given
by the eq.(2.6-2.14) with only inhomogeneous terms
proportional to $t_{\perp}^2$ kept
\begin{equation}
g_{i}(\xi)=C_{i}t_{\perp}^{2}(0){e^{2\Delta_{t_\perp}\xi}-e^{\Delta_{i}\xi}
\over{2\Delta_{t_\perp}-\Delta_{i}}}
\end{equation}
where $\Delta_{i}, i=1,...,9,BS,U$ denote dimensions of relevant
operators and $C_i$ are the coefficients
standing in front of terms
proportional to $t^{2}_{\perp}$ in the r.h.s. of (2.6-2.14).

It follows from (2.22)
that $g_1 (\xi_0)=-g_2 (\xi_0)=-g_3 (\xi_0)=g_4 (\xi_0)=-\lambda/2$
 while all the other couplings  $g_i(\xi_0),i=5,...,9$ are of
 order $\lambda^2$. On the other hand
at $\xi>\xi_0$ one can also omit in (2.6-2.14) all inhomogeneous terms
using $g(\xi_0)$ as bare values.
Naively, it could mean that one has to account the
 leading couplings $g_{1,2,3,4}$ plus $g_{BS}, g_U$ first and then to treat
all the rest as additional perturbations.
However it turns out that the solution is not so straightforward.

Let us consider first the case away from half filling.
Then it can be shown that  couplings $g_{2}, g_3$ and $g_{BS}$ all tend to zero
though $g_1$ diverges. Asymptotically the following relations
hold
\begin{eqnarray}
\frac{g_2 (\xi)}{g_{1}(\xi)}\sim \exp(\xi(K_{c}^{-}-{1\over K_{c}^{-}}
+{1\over K_{s}^{-}}-K_{s}^{-}))\rightarrow 0\nonumber\\
\frac{g_3 (\xi)}{g_1 (\xi)}\sim \exp(\xi(K_{c}^{-}-{1\over K_{c}^{-}}
+{1\over K_{s}^{-}}-K_{s}^{+}))
\rightarrow 0\nonumber\\
\frac{g_{BS} (\xi)}{g_1 (\xi)}\sim \exp(\xi(K_{c}^{-}+{1\over K_{s}^{-}}
-{K_{s}^{-}}-K_{s}^{+}))
\rightarrow 0\nonumber\\
{1\over K_{c}^{-}(\xi)}-{1\over K_{c}^{-}(0)}\sim {K_{s}^{-}(\xi)}\sim
{g_{1}^{2}(\xi)-g_{1}^{2}(0)\over {2-K_{c}^{-}(\xi)-{1\over K_{s}^{-}(\xi)}}}
\end{eqnarray}
Thus we infer that $K^{-}_{c}(\xi)$ vanishes while $K^{-}_{s}(\xi)$
goes to infinity. But this means that the assumption about smallness of
couplings $g_{5,...,9}$ made on the basis of their values at $\xi=\xi_0$
 was not quite correct.
Namely one has to include those terms which contain one of the fields
$\phi_{c}^{-}$ or $\theta_{s}^{-}$ which are "close" to get locked.
A simple inspection yields that the second relevant coupling (besides $g_1$)
is $g_6$ while in the case of $\rho=1$ one has to keep $g_{4,8,U}$ as well.
The resulting system of
equations in the range $\xi_0<\xi<{1\over \lambda}$ reads as
\begin{equation}
{{d g_{1}}\over {d\xi}}
={1\over 2}(z_{c}^{-}-z_{s}^{-})g_1-g_4 g_U
\end{equation}
\begin{equation}
{{d g_6}\over {d\xi}}
={1\over 2}(z_{s}^{-}-z_{c}^{-})g_6 -g_4 g_{8}
\end{equation}
\begin{equation}
{dz_{c}^{-}\over {d\xi}}
=g_{1}^{2}+g_{U}^{2}
\end{equation}
\begin{equation}
{{dz^{-}_{s}}\over {d\xi}}
=-g_{1}^{2}-g_{6}^{2}-g_{4}^{2}
\end{equation}
\begin{equation}
{{dz^{+}_{s}}\over {d\xi}}
=g_{6}^{2}+g_{8}^{2}
\end{equation}
\begin{equation}
{{dz^{+}_{c}}\over {d\xi}}
=g_{4}^{2}+g_{8}^{2}+g_{4}^{2}
\end{equation}
Away from half filling all $g_{4,8,U}$
freeze at $\xi\sim \ln{1\over
\delta}$ (and, consequently,  $z^{+}_{c}$ is frozen too)
 and there are only $g_{1,6}$ left over.
Then the  system (2.24-2.29)
demonstrates a development of the strong coupling regime in all channels
except the "+" charge one (namely, $g_{1}(\xi),g_{6}(\xi)\rightarrow {-\infty}$
and $z_{c}^{-}(\xi),z^{+}_{s}(\xi)\rightarrow\infty$ while $z^{-}_{s}(\xi)
\rightarrow{-\infty}$). As usual, these tendencies have to be undersood
in such a way that at $\xi\sim \frac{1}{\lambda}$ all couplings reach values
of order unity and don't vary further.

Including couplings  $g_{4,8,U}$ at $\delta\rightarrow 0$
 one can see that they don't alter
the behavior found for the doped case while  the
"+" charge sector is also driven to the
strong coupling regime in accordance with  the complete freezing of
charge degrees of freedom at $\rho=1$.

 To facilitate the analysis of leading instabilities of the
complete Hamiltonian
(2.3,2.5) one has to consider eight relevant order parameters
 where plus and minus correspond to intra- versus inter-chain
type of ordering
\begin{equation}
CDW_{+}=\sum\Psi^{f\dagger}_{\mu\sigma}\Psi^{f}_{-\mu,\sigma}\sim
\cos(\phi^{+}_{c}+\phi^{-}_{c})\cos(\phi^{+}_{s}+\phi^{-}_{s})\end{equation}
\begin{equation}
CDW_{-}=\sum\Psi^{f\dagger}_{\mu\sigma}\Psi^{-f}_{-\mu,\sigma}\sim
\cos(\phi^{+}_{c}+\theta^{-}_{c})
\cos(\phi^{+}_{s}+\theta^{-}_{s})\end{equation}
\begin{equation}
SDW_{+}=\sum \Psi^{f\dagger}_{\mu\sigma}\Psi^{f}_{-\mu,-\sigma}\sim
\cos(\phi^{+}_{c}+\phi^{-}_{c})\cos(\theta^{+}_{s}+\theta^{-}_{s})
\end{equation}
\begin{equation}
SDW_{-}=\sum \Psi^{f\dagger}_{\mu\sigma}\Psi^{-f}_{-\mu,-\sigma}\sim
\cos(\phi^{+}_{c}+\theta^{-}_{c})
\cos(\theta^{+}_{s}+\phi^{-}_{s})\end{equation}
\begin{equation}
SS_{+}=\sum\sigma\Psi^{f}_{\mu\sigma}\Psi^{f}_{-\mu,-\sigma}\sim
\cos(\theta^{+}_{c}+
\theta^{-}_{c})\sin(\phi^{+}_{s}+\phi^{-}_{s})\end{equation}
\begin{equation}
SS_{-}=\sum\sigma\Psi^{f}_{\mu\sigma}\Psi^{-f}_{-\mu,-\sigma}\sim
\cos(\theta^{+}_{c}+
\phi^{-}_{c})\sin(\phi^{+}_{s}+\theta^{-}_{s})\end{equation}
\begin{equation}
TS_{+}=\sum\sigma\Psi^{f}_{\mu\sigma}\Psi^{f}_{-\mu,\sigma}\sim
\cos(\theta^{+}_{c}+\theta^{-}_{c})\sin(\theta^{+}_{s}+\theta^{-}_{s})
\end{equation}
\begin{equation}
TS_{-}=\sum\sigma\Psi^{f}_{\mu\sigma}\Psi^{f}_{-\mu,\sigma}\sim
\cos(\theta^{+}_{c}+\phi^{-}_{c})\sin(\theta^{+}_{s}+\phi^{-}_{s})
\end{equation}
We remind that if any of the fields ($\phi_{c,s}^{\pm}$
or $\theta_{c,s}^{\pm}$)
gets locked then the corresponding cosine
acquires a nonzero expectation value
and  $<\cos\phi(x)\cos\phi(0)>\rightarrow |<\cos\phi(0)>|^2$
as $x$ tends to infinity.
On the other hand, fluctuations of both this variable
and its dual one become gapful. Formally one can identify the state where
$\phi_{c,s}^{\pm}$ is ordered with the limit $K_{c,s}^{\pm}\rightarrow 0$
while $\theta_{c,s}^{\pm}$ becomes ordered at $K_{c,s}^{\pm}
\rightarrow \infty$.

Then one can easily see that
in the case when $\theta_{s}^{-}$ and $\phi_{s}^{+}$ are locked
the only competing instabilities  are the interchain charge density wave
corresponding
to the order parameter
$CDW_{-}\sim\cos(\phi^{+}_{c}+
\theta^{-}_{c})$ and the
interchain singlet pairing
described by $SS_{-}\sim\cos(\theta^{+}_{c}+\phi^{-}_{c})$.
In fact, the former state  can be also recognized
as a counterpart of the two-dimensional
flux phase. Indeed, this state is characterised
by the  commensurate with density flux $\Phi=2k_{F}$ which is
defined as a circulation
of a phase of the on-rung order parameter
 $<u^{\dagger}_{i}d_{i}+d^{\dagger}_{i}u_{i}>$  through a plaquette
formed by two adjacent
rungs of the ladder. In the case of spinless fermions this type of ordering
called "Orbital Antiferromagnet"
was first  considered in \cite{N} as a prototype of recently
proposed two-dimensional flux states.

Although the flux phase can be in principal realized in some extended models
we see that in our case of the double chain Hubbard model
where the field
 $\phi_{c}^{-}$ also gets
locked the ground state is a spin gapped singlet superconductor.

It is also instructive
to express the above order parameters in terms of the hybridized one-particle
states corresponding to the abovementioned "bonding" and "antibonding"  bands
\begin{eqnarray}
CDW_{-}=\sum_{\sigma}
A^{\dagger}_{R\sigma}A_{L\sigma}-B^{\dagger}_{R\sigma}B_{L\sigma}\nonumber\\
SS_{-}=\sum_{\sigma}
A^{\dagger}_{R\sigma}A^{\dagger}_{L,-\sigma}
-B^{\dagger}_{R\sigma}B_{L,-\sigma}^{\dagger}
\end{eqnarray}
Considering the distribution of signs of the order parameter
 $SS_{-}$ on the "four-point Fermi surface"
$({\vec k}=(k_F ,0),(-k_F ,0),(k_F ,\pi),(-k_F ,\pi))$
we observe that it corresponds to the "d-wave" type pairing.
We conjecture
that in a
two-dimensional array of weakly coupled double chains with a continuum
 Fermi surface this
type of ordering does transform into an ordinary d-wave pairing.

\section{Conclusions}
In the present paper we applied the bosonization method
 to find further arguments in support
of the recently proposed scenario of singlet superconductivity in the spin gap
state of doubled
Luttinger chains. Previous results obtained in the framework of
the mean field approach \cite{SRZ} as well as earlier numerical studies
\cite{DRS},\cite{NWS}  also testify in favor of this picture.

We also want to stress that  our conclusions contradict
with a recent claim about
the existence of the
strong coupling fixed point where some spin excitations remain
gapless made
in \cite{ZFL}. These authors considered the double chain $t-J$ model
without an interchain spin exchange ($J_\perp =0$).
Then on the bare level  their Hamiltonian can be assigned to the universality
class of the purely
forward scattering  model considered in \cite{FL}.
In this special
case indeed the only
field
becoming massive
is $\theta^{-}_{s}$. In principal, it can't be ruled out that for some specific
double chain models only part of all  relevant fields acquire masses and the
others remain massless. One example of this type was discussed by the
authors of \cite{SBi} who found only $\phi_{c}^{-}$ and $\phi_{s}^{+}$
to be massive in the framework of the model including solely an interchain
interaction of fermions with opposite spins.

However our investigation of the Hubbard-type models shows that
the presence of the interchain
one particle hopping is already sufficient to generate
the
antiferromagnetic spin exchange term
with $J_\perp \sim \frac{t^{2}_{\perp}}{max(t,U)}$
which makes all spin modes gapful at $\rho\lsim 1$.
 We believe that spin liquid behavior
with a finite spin gap which evolves into "d-wave" pairing
upon doping is a robust feature
of a whole
variety of isotropic spin
 models of weakly coupled and strongly correlated fermions on double chains.

\section{Acknowledgements}
The authors are grateful to Profs. P.B.Wiegmann and F.D.M.Haldane for
valuable discussions.
This work was supported by the Swiss National
Fund. One of the authors (D.V.K.) also acknowledges
the support from the NSF Grant.

\pagebreak

\end{document}